\documentclass{elsart}

 \usepackage{epsfig}

\usepackage{amssymb}

\begin{document}

\begin{flushright}

{\bf DFUB 6/2003}

\end{flushright}
\begin{frontmatter}

 \title{Monte Carlo simulation of an experiment
 looking for radiative solar neutrino decays}

 \author[bof,tesre]{S. Cecchini}
 \author[bof]{D. Centomo}
 \author[bof]{G. Giacomelli}
 \author[bof,iss]{V. Popa}
 \author[bof,iss]{C.G. \c{S}erb\u{a}nu\c{t}}
\address[bof]{Dipartimento di Fisica dell'Universit\`{a} and INFN Sezione di Bologna,
 I-40127 Bologna, Italy}
\address[tesre]{IASF/CNR, I-40129 Bologna, Italy}
\address[iss]{Institute for Space Sciences, R-77125 Bucharest M\u{a}gurele,
Romania}

\begin{abstract}
We analyse the possibility of detecting visible photons from a hypothetical radiative
decay of solar neutrinos. Our study is focused on the simulation of such
measurements during total solar eclipses and it is based on the BP2000 Standard Solar
Model and on the most recent experimental information
 concerning the neutrino properties. Our calculations yield the
 probabilities of the decays, the shapes of the visible signals and the spectral
 distributions of the expected photons, under the assumption that solar neutrino
 oscillations occur according to the LMA model.
\end{abstract}

\end{frontmatter}

\section{Introduction}
\label{intro}

In the last few years it has become  clear that neutrinos have non-vanishing
masses, and that the neutrino flavor eigenstates ($\nu_e$, $\nu_\mu$ and $\nu_\tau$)
are superpositions of mass eigenstates ($\nu_1$, $\nu_2$ and $\nu_3$). For a recent
review, see \cite{gg}. In this context, neutrinos could
undergo radiative decays, e.g. $\nu_2 \rightarrow
\nu_1 + \gamma$, as initially suggested in \cite{sciama}. The present
status of decaying theory is sumarized in \cite{sciama2}.
Such decays request
that the involved neutrinos have a non-vanishing  electric dipole moment; the
very stringent existing experimental limits refer to the flavor neutrino eigenstates
and they are not directly applicable to possible dipole moments of mass neutrino
eigenstates.

In a pioneering experiment  performed during the
Total Solar Eclipse (TSE) of October 24, 1995
a search was made for visible photons emitted through possible radiative decays
of solar neutrinos during their flight between the Moon and the Earth \cite{vanucci}.
 In the analysis
  of the data, the authors assumed
  that all neutrinos  have masses of the order of
few eV,   $\Delta m_{12}^2 = m_2^2 - m_1^2
 \simeq 10^{-5}$ eV$^2$, and an average neutrino energy of
860 keV; furthermore they assumed that all decays would lead to visible photons,
which would travel nearly in the same direction as the parent neutrinos, thus
leading to a narrow spot of  light
 coming from the direction of
 the center of the dark disk of the
Moon.

Subsequently,
Fr\`{e}re and Monderen made more accurate calculations on the shape of the expected
signal \cite{frere}, considering the Sun as an extended  source of electron
neutrinos with
 typical energies of the order of 1 MeV,
a Lagrangian formalism for the radiative decay,
  different neutrino masses (1 eV or 0.5 eV), and mass
  squared
  differences $\Delta m^2$ of
  $10^{-5}$, 0.25 and 1 eV$^2$; they have  shown
that the expected signal could have an extended angular pattern.

Some of the authors of this paper
 were involved in two experiments along the line of \cite{vanucci}, during
the total solar eclipses of August 11, 1999 (in Romania)
\cite{n1,n2,n3}, and of June 21, 2001 (in Zambia) \cite{n3}.
In 1999 the bad weather conditions did not allow the planned observations, but
we could use a videotape filmed by a local television (R\^{a}mnicu V\^{a}lcea).
 The analysis of
the data was performed
in the hypothesis of a possible
\begin{equation}
\label{deca}
\nu_2 \rightarrow \nu_1 + \gamma
\end{equation}
decay, with $m_2 > m_1$.
 For the analysis of the 1999 data we have chosen
 the two $\Delta m^2$ values suggested by the
MSW SMA (Small Mixing Angle) and LMA (Large Mixing Angle) solutions of the Solar
Neutrino Problem (SNP), allowed by the then available experimental data from solar
neutrino experiments.
We developed a Monte Carlo (MC) simulation of the radiative
solar neutrino decay, considering the solar neutrino energy spectrum predicted
by the Standard Solar Model (SSM) \cite{bahcall1} and the mass of the $\nu_1$
in the range of 1 - 10 eV, as it was expected at that time.
 Since the angular resolution of the data was not very
good,
we considered the Sun as a pointlike neutrino surce. The simulation has shown that
the expected signal should be a narrow spot of light in the direction of the center
of the Sun, and allowed an evaluation of the fraction of decays yielding visible
photons as function of the chosen neutrino $\nu_1$ mass $m_1$
and $\Delta m^2$ values.

The 2001 experiment lead to better quality data, so the real
spatial distribution of
the solar neutrino yield had to be considered. Furthermore, the recent SNO
results \cite{sno1,sno2}  favour  the LMA solution  and
could indicate also the presence of a $\nu_\tau$ component in the solar neutrino
flux at the Earth level.

The WMAP (Wilkinson Microwave Anisotropy Probe)
results after the first year of flight
\cite{map} limit the sum of the masses of the three
neutrino species to 0.23 eV ( 95\%
Confidence Level).

In this paper we present a more complete simulation.
We assume that
$m_1 < m_2 < m_3$ where $m_1$, $m_2$, $m_3$ are the masses
of the $\nu_1$, $\nu_2$ and
$\nu_3$ mass eigenstates, respectively; but we
restrict our analysis to a two generation mixing
scenario, assuming the present
mass differences obtained from solar neutrino experiments,
the LMA solution with $\Delta m^2_{12} = 6 \times 10^{-5}$ eV$^2$.
Since SNO suggests also
the presence of $\nu_3$ in the solar neutrino flux, we considered also the mass
difference measured by atmospheric neutrino experiments \cite{macro,miri,sk}:
$\Delta m^2_{13} \simeq
\Delta m^2_{23} = 2.5 \times 10^{-3}$ eV$^2$. We included in our MC
all the details of neutrino production in the Sun, as given by the
``BP2000" SSM \cite{bahcall2}.

The aim of our simulation is to give information on the shape of a possible decay
signal of solar neutrinos (angular and energy  distributions
 of the emitted photons) and to estimate
the probabilities for the decay and the geometrical detection efficiency.
This information should allow to obtain limits on
the neutrino lifetimes from the experimental observations. As it
is shown in the following sections, some of the input parameters of
the code refer to the characteristics of a specific experiment. The results
presented in this paper are obtained in the conditions of the observations
made in 2001 in Zambia \cite{n3} with a digital videocamera.

\section{The Monte Carlo simulation}
\label{mc}

The  assumed geometry for the simulation of the radiative decay of solar
neutrinos during a TSE is shown in Fig. 1.
The notations in the figure will
be used in the following subsections.

\begin{figure}
\vspace{-10mm}
\begin{center}
       \mbox{  \epsfysize=9cm
               \epsffile{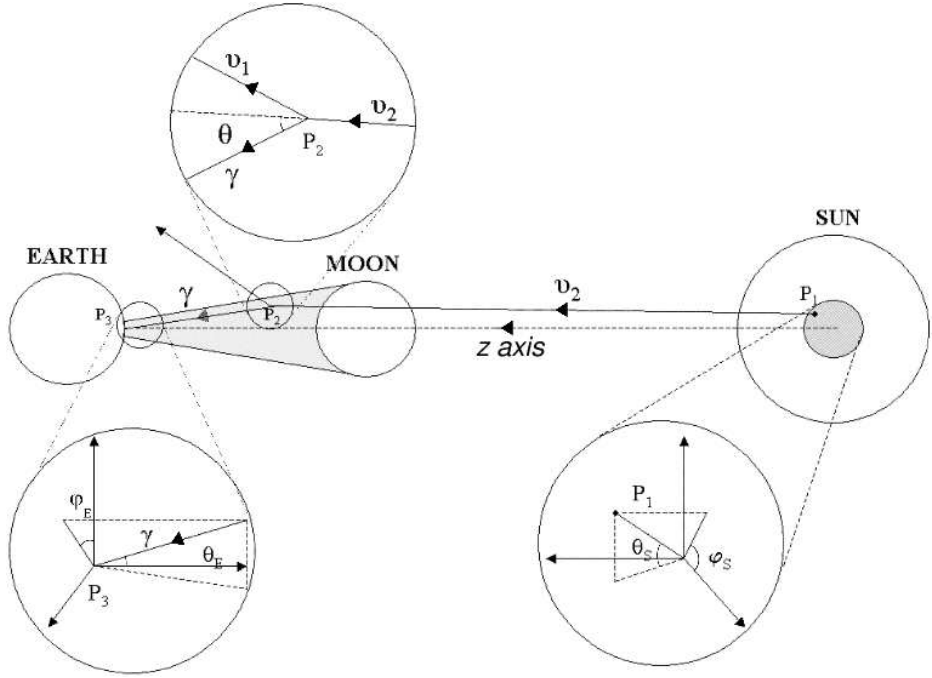}}
\caption{A sketch of the geometry of the production
of solar neutrinos, their possible radiative decay (in the space between the
Moon and the Earth) and the detection of the emitted photon, during a TSE.
The $z$ axis is directed from the center of the Sun to the observation point
on the Earth (or center of the Earth)}
\end{center}
\end{figure}

\subsection{Neutrino production inside the Sun}
\label{soare}

In simulating the solar neutrino production, we used the ``BP2000" SSM \cite{bahcall2}
in its numerical form directly available from \cite{web}.
The first step consists in chosing a specific reaction/decay
 yielding  neutrinos (both
from the p-p and the CNO cycles) according to its predicted contribution
to the solar neutrino flux at the Earth. The neutrino energy and
the radius $R$ of its production point are then randomly generated according
to the SSM. In Fig. 2
 we present the neutrino solar energy spectrum
obtained from
$3 \times 10^7$ generated neutrinos, and the
distribution of the distance $R$ from the center of the Sun of the production points.

\begin{figure}
\begin{center}
       \mbox{  \epsfysize=7cm
               \epsffile{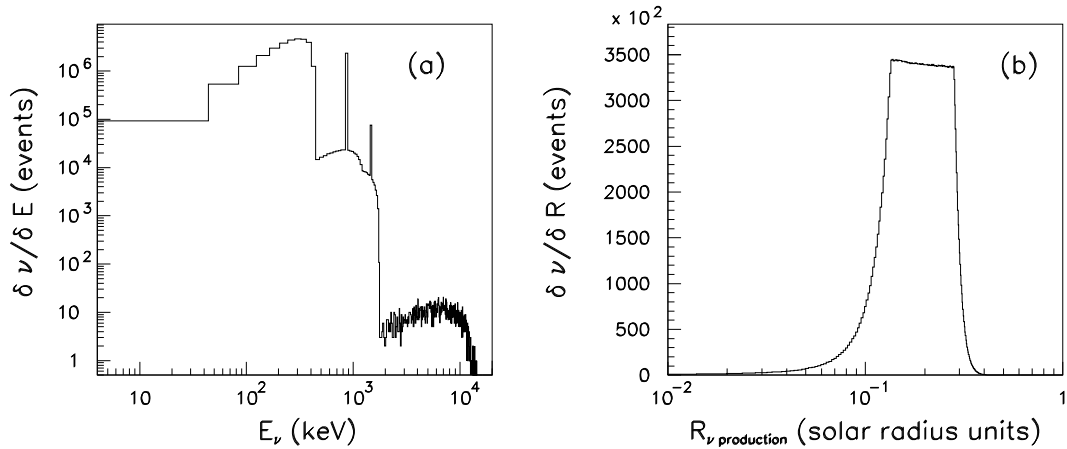}}
\caption{Monte Carlo simulations according to the ``BP2000" SSM. (a)The energy
spectrum of solar neutrinos; the $^8B$ neutrinos are the few neutrinos with energy
between $10^3$ and $10^4$ keV.
(b) The distribution of the radial distance of the production points from the center of
the Sun.
}
\end{center}
\end{figure}

In order to ensure a uniform distribution of the birth points of the neutrinos
($P_1$ in Fig. 1) on each shell $\Delta R$ at radial distance $R$,
we determine for each neutrino
the angular spherical coordinates
 generating uniformely the cosinus of the zenithal angle ($\theta_S$ in Fig. 1,
assuming the ``z" axis
oriented from the centre of the Sun towards the centre of the Earth) and the
polar angle ($\phi_S$ in Fig. 1.).

\subsection{Propagation of solar neutrinos and of the decay photons}
\label{pamant}

The solar neutrinos are produced as $\nu_e$'s, thus as a superposition
of neutrino mass states. After leaving the Sun, during their flight in the
interplanetary space, their weak flavor is irrelevant.

In our simulation we
consider the $\nu_2$ (or $\nu_3$) component of the neutrino flux, that may decay into
the lower mass state $\nu_1$ and a photon. As we are interested in the possibility
to observe such photons during a total solar eclipse, we impose that the decay
processes take place in the space between the Moon and the Earth,
inside the shadow cone of the Moon (otherwise the separation of the decay photons
from the solar light backround would be impossible). Furthermore,
the decay photons must reach the detector (a CCD-equipped telescope,
a digital camera
or some other observation systems of the same kind).

The next step in the simulation is to define the incidence direction of the decay
photon on the detector, situated on the Earth surface at the location $P_3$
(see Fig.~1). We generate uniformly the cosinus value of the ``local" zenith
angle $\theta_E$ inside the shadow cone (or in the cone subtended by the
telescope or by the analysis apperture if smaller)
and then the value of the ``local" azimuthal
angle $\phi_E$, as defined in Fig.~1.

The probability of neutrino radiative decays (thus of their lifetime) depends on
their electric dipole momentum, which, for the weak flavor eigenstates, is
 limited by the existing experimental data \cite{pdg}.
Those limits are not directly applicable for the mass eigenstates. We may  assume that
the lifetime of $\nu_2$ is much larger than
the time of flight from the Sun to the Earth (otherwise the experiment would
be impossible). This implies that the decay points of the massive solar neutrinos
are uniformely distributed along their path from the Moon to the
Earth.

As we already defined the
direction of the ``detected" photon, we cannot choose the decay point of the neutrino
using the above assumption; instead we take advantage of the nearly cylindrical
symmetry of our problem to choose
the decay point (point $P_3$ in Fig. 1)
randomly along the photon
path and then determine the neutrino path as the segment starting in $P_1$ and
ending in $P_3$. This does not affect the expected uniformity of the distribution
of the decay points along the neutrino path from the Moon to the Earth, as shown
in Fig.~3a. We also checked  the isotropy on the incidence of
solar neutrinos on the Moon surface (see Fig. 3b).

\begin{figure}
\vspace{-17mm}
\begin{center}
       \mbox{  \epsfysize=7cm
               \epsffile{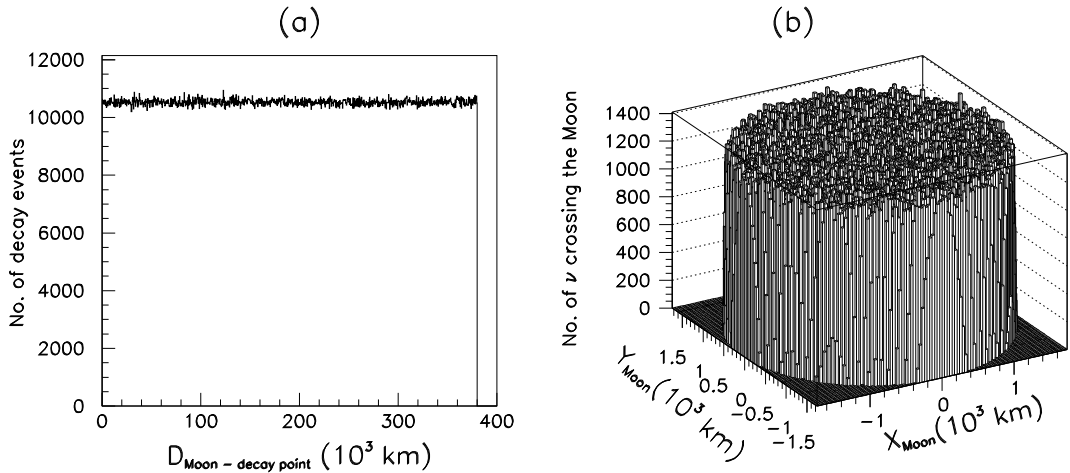}}
\caption{``Isotropy" tests for the simulated solar neutrino flux:
a) The distribution of neutrino path lengths from the Moon till their
decay point. b) The distribution of the points of incidence of the solar $\nu$'s
on the Moon disk.
One may notice that both distributions are uniform}
\end{center}
\end{figure}

\subsection{Neutrino decays}
\label{decay}

For every simulated event we know (from the steps described
in the above subsections) the 4-vector of the ``heavy" neutrino $\nu_2$, its
birth and decay points, and the direction of the photon emission;
from the 4-momentum conservation we obtain the energy (in the
laboratory reference frame, LRF) of the emitted $\gamma$
\begin{equation}
E_\gamma = \frac{\Delta m^2}{2}~\frac{1}{E_\nu-p_\nu \cos\theta}~,
\end{equation}
where $E_\nu$ and $p_\nu$ are the energy and the momentum of the parent $\nu_2$
neutrino, respectively, and $\theta$ is the emission angle of the photon relative to
the direction of the $\nu_2$ momentum in the LRF (see Fig. 1).

In the center of mass (CM) of the decaying $\nu_2$ neutrino, the probability
density of the emission of a photon at the zenithal angle $\theta^*$ is given by:
\begin{equation}
\frac{d\Gamma}{d(\cos \theta^*)}=K(1+\alpha \cos \theta^*)
\end{equation}
where the $\alpha$ parameter depends on the polarization state of the initial
$\nu_2$ flux: $\alpha = 0$ for unpolarized (Majorana) neutrinos, and
$\alpha = \mp 1$ for left and right handed (Dirac) $\nu_2$'s, respectively.
The constant $K$ comes from the general description of the decay of a fermion into
a fermion and a boson
\begin{equation}
K = \frac{\alpha_e^2}{\pi^2} \frac{m_2}{(\Delta m^2)^3}(m_1^2 + m_2^2 +
m_1 m_2),
\end{equation}
where $\alpha_e$ is the fine structure constant.

In order to estimate the probability of each simulated event,
one would have to integrate Eq. 3 inside the solid angle under which the
emitted photon ``sees" the detector. This is unpractical since the physical
dimension of any usable detector is too small compared to the distance scales
involved in the simulation.
 An equivalent approach is based on the observation that any
imaging system (such as a CCD in the focal plane of some optical system) has a limited
angular resolution, as the images of all point sources inside the solid angle covered
by a single pixel will be superimposed. We can then numerically integrate Eq. 3
on the area around the decay point $P_2$ (see Fig. 4 for a
sketch; the integration area is represented as the dark square centered in $P_2$)
 that corresponds
to the angular aperture of a single pixel. The probabilities obtained
in this way are used as weights for the MC generated events.

\begin{figure}
\vspace{-10mm}
\begin{center}
       \mbox{  \epsfysize=8cm
               \epsffile{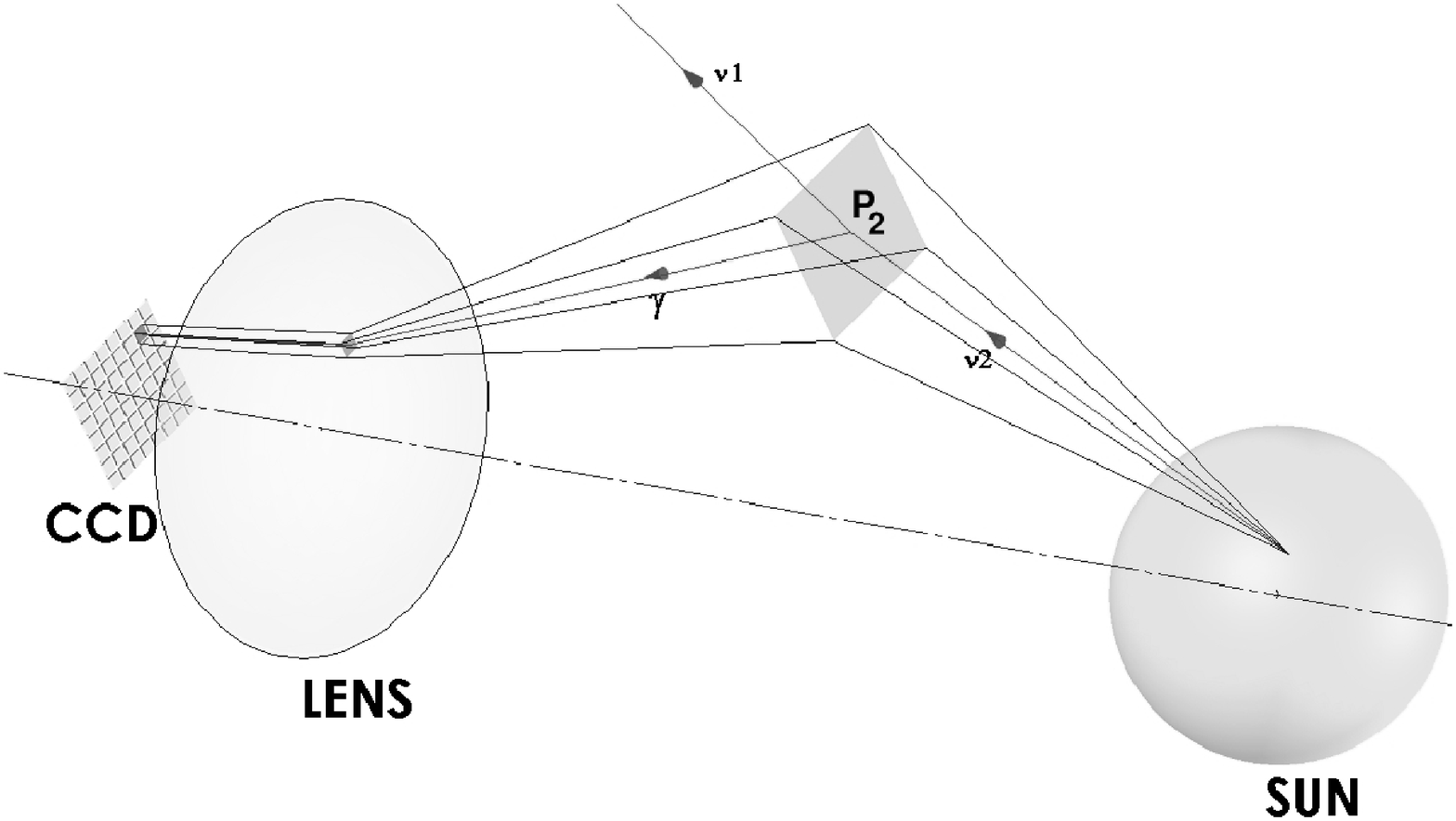}}
\caption{A sketch showing the integration area for the probability density in Eq. 3.
Any decay event inside the dark square around point $P_2$ would lead to a signal
in the same CCD pixel. (In this sketch the Moon is not represented.)}
\end{center}
\end{figure}

\section{Results and discussions}

The results presented in this  Section  are obtained assuming
an angular resolution of each pixel of 10", and an angular aperture
of the analysis (the maximum value of $\theta_E$ in Fig. 1) of 480",
as in the case of the digital
video-camera used during the 2001 TSE \cite{n3}. The mass $m_1$ of the $\nu_1$
mass eigenstate was considered in a range between $10^{-3}$ eV and 0.3 eV; MC
simulations
were made for $\Delta m^2 = 6 \times 10^{-5}$ eV$^2$ and also for $\Delta
m^2 = 2.5 \times
10^{-3}$ eV$^2$. For the polarization parameter $\alpha$ in Eq. 3, we considered three
possible values: -1, 0 and 1.

For each combination of neutrino mass and $\Delta m^2$ we
requested $3 \times 10^4$ unweighted events yielding photons in the visible range.
Assuming $\Delta m^2 = 6 \times 10^{-5}$ eV$^2$ the total number of iterations was
about $1.6 \times 10^9$, while for $\Delta m^2 = 2.5 \times 10^{-3}$ eV$^2$, about
$3.3 \times 10^7$ generated events were needed. The simulated events were then weighted
according to 
the integral of Eq. 3, as discussed in the previous Section.
In weighting the events the factor $K$ in Eq. 3 was set to 1, as it is a constant
for each $(m_2,~\Delta m^2)$ hypothesis.
For this
analysis we also imposed the initialization of the random number
generator to be the same for each run.

Fig. 5 shows the fraction of visible photons, versus
the mass of the lighter neutrino $\nu_1$. Fig. 5a corresponds to
$\Delta m^2 = 6 \times 10^{-5}$ eV$^2$, while the results for $\Delta m^2 = 2.5 \times
10^{-3}$ eV$^2$ are presented in Fig. 5b. The light triangles are obtained
assuming $\alpha = 1$, while the dark and light circles correspond to $\alpha = 0$
and $\alpha = +1$, respectively.

\begin{figure}
\vspace{-65mm}
\begin{center}
       \mbox{  \epsfysize=14cm
               \epsffile{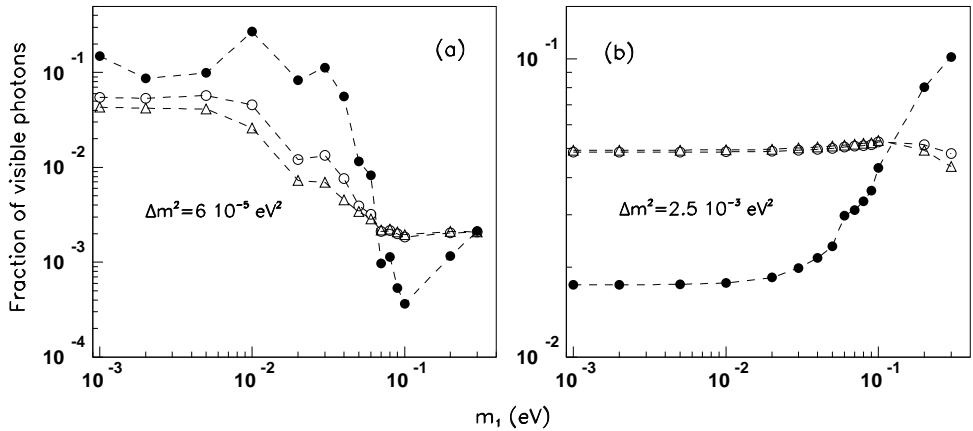}}
\caption{The fraction of visible photons produced in the simulated radiative decays
of massive solar neutrinos. The assumed squared mass differences are (a) $\Delta
m^2 = 6 \times
10^{-5}$ eV$^2$ and (b) $2.5 \times 10^{-3}$ eV$^2$ (right). The light triangles,
 light circles and  dark circles correspond to polarisations $\alpha = -1$,
0 and +1, respectively. The dashed lines are drawn only to guide the eye.}
\end{center}
\end{figure}

As discussed in the previous Section, for each simulated neutrino decay event we
numericaly integrated Eq. 3, thus obtaining a probability that includes the
contributions from the kinematics of the decay itself as well as from the
{\em a priori} request included in the simulation that the emitted photon
reaches the detector. In Fig. 6 we present these probabilities, averaged over
all Monte Carlo events yielding visible photons,
 versus the mass of the $\nu_1$ neutrino.
Fig. 6a refers to $\Delta m^2 = 6 \times 10^{-5}$ eV$^2$, and Fig. 6b
to $\Delta m^2 = 2.5 \times 10^{-3}$ eV$^2$. The symbols used for different
polarization states are the same as in Fig. 5.

\begin{figure}
\vspace{-65mm}
\begin{center}
       \mbox{  \epsfysize=14cm
               \epsffile{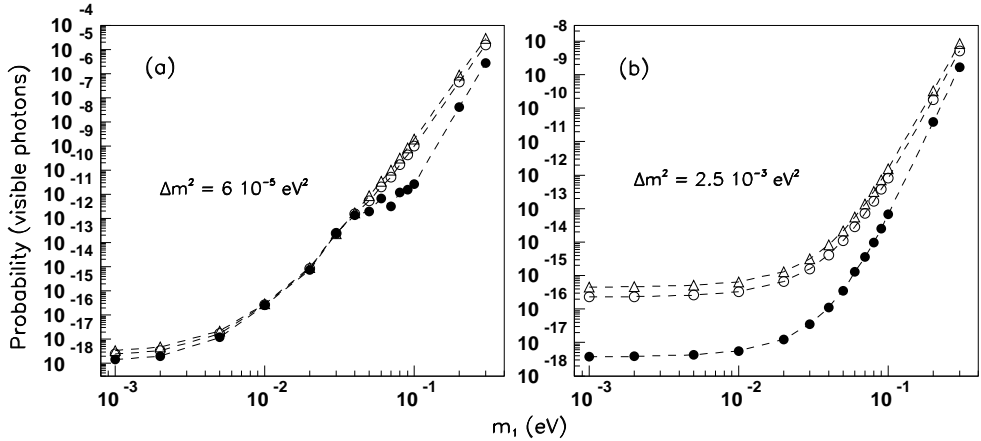}}
\caption{ Average probabilities for the neutrino radiative decay yielding
visible photons inside the simulated analysis acceptance.
 The assumed squared mass differences are $\Delta
m^2 = 6 \times
10^{-5}$ eV$^2$ (a) and $2.5 \times 10^{-3}$ eV$^2$ (b). The light triangles,
the light circles and the dark circles correspond to polarisations $\alpha = -1$,
0 and +1, respectively. The dashed lines are only meant to guide the eye.}
\end{center}
\end{figure}

The probabilities shown in Fig. 6 are very small, but one should consider that
they apply to all the solar massive neutrinos that cross the ``acceptance cone"
of the detector between the Earth and the Moon (see Fig. 1).

Both Figs. 5 and 6 show a strongly non-linear behaviour, as the condition
imposed to the photons to be in the visible spectrum selects different regions
of the solar neutrino spectrum for different mass or polarization hypothesis. This
effect is illustrated in Fig. 7, for three values of the neutrino mass $m_1$,
0.001, 0.01 and 0.1 eV: the solid, dashed and dotted histograms respectively.
Fig. 7a corresponds to $\Delta m^2 = 6 \times 10^{-5}$ eV$^2$, while
Fig. 7b to $\Delta m^2 = 2.5 \times 10^{-3}$ eV$^2$. In all cases the polarization
parameter $\alpha$ was assumed -1.

\begin{figure}
\vspace{-20mm}
\begin{center}
       \mbox{  \epsfysize=8cm
               \epsffile{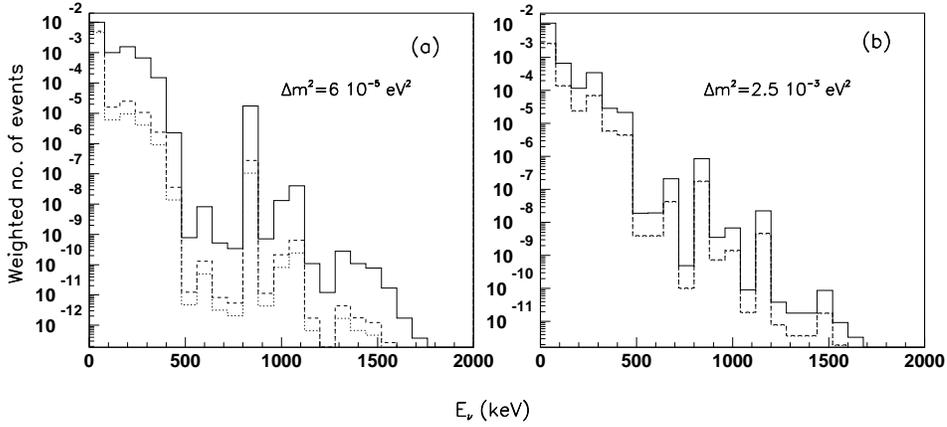}}
\caption{The energy distribution of the solar neutrinos that yield visible photons
through radiative decay, assuming $m_1 = 0.001$ eV (solid histograms), 0.01 eV
(dashed histograms) and 0.1 eV (dotted histograms). The squared mass difference
is assumed $6 \times 10^{-5}$ eV$^2$ (a) and $2.5 \times 10^{-3}$ eV$^2$ (b).
In all cases $\alpha = -1$.}
\end{center}
\end{figure}

 Fig. 7 suggests that the high energy solar $^8B$ neutrino tail does not
yield  visible photons through radiative decays. Such process
could instead happen for low energy $pp$ neutrinos, with some contributions
from the $^{13}N$, $^{15}O$ neutrinos and from the $^7Be$ and $pep$ lines;
notice that these last contributions are more
noticeable in Fig. 7a than in Fig. 7b.

In conducting an experiment searching for visible photons from a hypothetical radiative
solar neutrino decay during TSE's and in choosing the appropriate data analysis
methodology, the simulation of the expected signal is very important. Fig. 8 presents
such simulations, in the same conditions and with the same conventions as in Fig. 7.
The shape of the signals corresponding to different neutrino masses and mass
differences (assuming left handed neutrinos, thus $\alpha = -1$) is presented as
the radial distribution of the ``detected" luminosity. This is equivalent to
the distribution of the weighted number of visible photons, averaged on
circular corona around the same value of the incidence zenith angle $\theta_E$
(see Fig. 1).

\begin{figure}
\begin{center}
       \mbox{  \epsfysize=8cm
               \epsffile{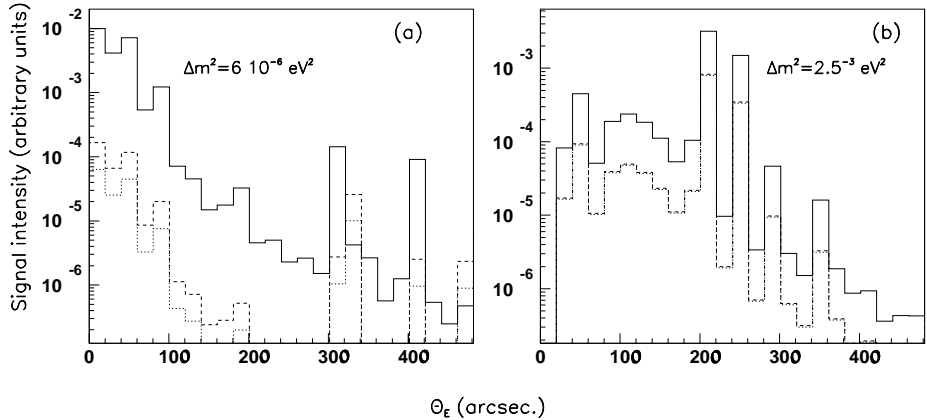}}
\caption{The expected shapes of the visible signals produced by the
hypothesized solar neutrino radiative decay,
 assuming $m_1 = 0.001$ eV (solid histograms), 0.01 eV
(dashed) histograms) and 0.1 eV (dotted histograms). The squared mass difference
is assumed to be $6 \times 10^{-5}$ eV$^2$ (a) and $2.5 \times 10^{-3}$ eV$^2$ (b).
In all cases $\alpha = -1$.}
\end{center}
\end{figure}

The histograms in Fig. 8a correspond to the simulated $\nu_2 \rightarrow \nu_1 +
\gamma$ decays. For all neutrino masses, the expected signal is concentrated at
small $\theta_E$ angles (about 50 arcsec).
The widths and shapes of the signals are
sensitive to the mass assumed: the larger the mass, the narrower the signal band.
 The two peaks seen at about 300 and 400 arcsec. for $m_1 = 10^{-3} eV$ could be
correlated with the contribution of $^7Be$ or $pep$ neutrinos; they are about
a factor 100 lower than the central maxima, so their contribution
is not measurable.

In the case of $\nu_3 \rightarrow \nu_1 + \gamma$ simulated decays (Fig. 8b),
the signal is
broader (about 250 arcsec) and is less sensitive to the mass choice.
The peaks observed at about
250" could have a similar origin as those in Fig. 8a, but could also be statistical
fluctuations.

If an experiment
on solar neutrinos
 has also a good energy resolution, then further information could
be obtained from the analysis of the spectra of the observed signal. Such spectra
are shown in Fig. 9, assuming only left-handed neutrinos, and considering the same
examples as in Figs. 7 and 8.

\begin{figure}
\begin{center}
       \mbox{  \epsfysize=8cm
               \epsffile{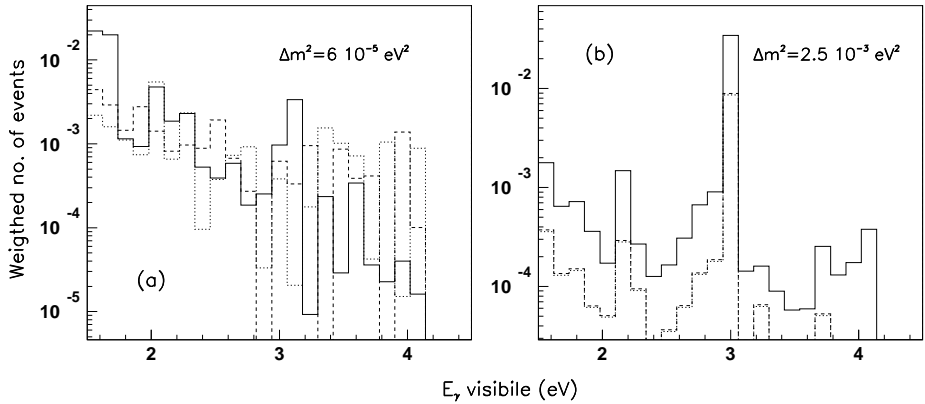}}
\caption{The expected energy spectra of the visible signals produced by the
hypothesized solar neutrino radiative decay,
 assuming $m_1 = 0.001$ eV (solid histograms), 0.01 eV
(dashed histograms) and 0.1 eV (dotted histograms). The squared mass difference
is assumed $6 \times 10^{-5}$ eV$^2$ (a) and $2.5 \times 10^{-3}$ eV$^2$ (b).
In all cases $\alpha = -1$.}
\end{center}
\end{figure}

As for the shape of the signal, its spectral decomposition seems to be more
sensitive to the neutrino mass values for the $\nu_2 \rightarrow \nu_1 +
\gamma$ decays (Fig. 9a). In this case most of the visible photons
are ``detected" in the red part of the spectrum, while the spectra in Fig.
9b suggest a dominant signal in the green, due to the larger mass difference
between $\nu_3$ and $\nu_1$.

Let us assume that an experiment as those simulated in this work would measure,
during a TSE, an excess of visible photons $\Phi_{obs.}$ from the
direction of the center of the sun, or, if the
search yields a negative result,
$\Phi_{obs.}$ is the photon flux of the observed fluctuations. In the first case,
estimates of the neutrino lifetime could be done; otherwise, a lower experimental
limit could be deduced.
Assuming that solar (electron) neutrinos are superpositions of only two mass
eigenstates,
\begin{equation}
|\nu_e > = |\nu_1> \cos \theta + |\nu_2> \sin \theta,
\end{equation}
where $m_2 > m_1$ and $\theta$ is the mixing angle, the average lifetime (or its
lower limit) $\tau$ of the $\nu_2$ neutrino could be computed from
\begin{equation}
N_\gamma = P \Phi_2 S_M t_{obs} \left(1-e^{-\frac{<t_{ME}>}{\tau}} \right)
e^{-\frac{t_{SM}}{\tau}},
\end{equation}
where $N_\gamma$ is the number of decay visible photons observed,
 $P$ are the mass - dependent probabilities shown in
 Fig. 6 a, $\Phi_2 = \Phi_\nu \sin^2 \theta$,(where $\Phi_\nu$ is the
flux of solar neutrinos at the Earth (or Moon) level), $S_M$ is the area of the
Moon surface covered by the analysis (the base of the cone of angle $\theta_E$
in Fig. 1) and $t_{obs}$ is the time of observation. $<t_{ME}>$ is the average
time spent by solar neutrinos inside the observation cone
(about one third of the flight
time from the Moon to the Earth), and $t_{SM}$ is the time of flight of the neutrinos
from the Sun to the Moon. The  low numerical values of the probabilities $P$
are compensated by the  large solar neutrino flux, combined with the large
area $S_M$, so an experiment as the one simulated here could yield at least
significant upper limits on the $\nu_2$ lifetime $\tau$.
 In the conditions of a 3.5 minutes long TSE (as that of 2001) observed with an
instrument with characteristics similar to those considered in this simulation,
one would expect a $\nu_2$ lifetime (in the proper reference frame)
sensitivity  ranging from few seconds to about $10^4$ seconds, assuming
$\nu_2$ neutrino masses of few $10^{-2}$ eV.

\section{Conclusions}

In this paper we presented a  Monte Carlo simulation for an experiment looking
for visible photons emitted by
 a possible solar neutrino radiative decay, during a total solar
eclipse. It was shown that for neutrino masses smaller than 0.1
eV and assuming squared mass differences in agreement with the Large Mixing
Angle Solution (LMA)
of the solar neutrino oscillations
\cite{sk}, such an experiment could give also an estimate
of the neutrino mass.

The analysis of the experimental data collected by some of the authors during the
2001 TSE in Zambia is being completed using the simulation results obtained
in this paper.

\section{Acknowledgements}

We would like to aknowledge many colleagues for useful comments and discussions.

This work was founded by NATO Grant PST.CLG.977691 and partially supported by the
Italian Space Agency (ASI) and the Romanian Space Agency (ROSA).


\begin{thebibliography}{00}




\bibitem{gg} G. Giacomelli and M. Sioli, Astroparticle Physics, hep-ex/0211035
\bibitem{sciama} A.L. Melott, D.W. Sciama,
{\em Phys.Rev.Lett.} {\bf 46} (1981) 1369-1372.
\bibitem{sciama2} D.W. Sciama, {\em Nucl.Phys.Proc.Suppl.} {\bf38} (1995) 320-323.
\bibitem{vanucci} C. Birnbaun et al., {\em Phys. Lett.} {\bf B397} (1997) 143-146.
\bibitem{frere} J.-M. Fr\`{e}re and D. Monderen, {\em Phys. Lett.} {\bf B431} (1998) 368-373.
\bibitem{n1} S. Cecchini et al. (NOTTE Coll.),
  {\em Astrophys. and Space Sci.} {\bf 273} (2000) 35-41.
\bibitem{n2} S. Cecchini et al. (NOTTE Coll.),
Limits on radiative decays of solar neutrinos
from a measurement during a solar eclipse, hep-ex/0011048
\bibitem{n3} V. Popa et al. (NOTTE Coll.),
 {\em Astrophys. and Space
Sci.} {\bf 282} (2002) 235-244.
\bibitem{bahcall1} J.N. Bahcall, Standard Solar Models, astro-ph/9808162
\bibitem{sno1} Q.R. Ahmad et al. (SNO Coll.), {\em Phys. Rev. Lett.} {\bf 87}
 (2001) 071301.
 \bibitem{sno2} Q.R. Ahmad et al. (SNO Coll.), {\em Phys. Rev. Lett.} {\bf 89}
 (2002) 011301.
\bibitem{map} D.N. Spergel at al., First year Wilkinson Microwave Anisotropy
Probe (WMAP) observations: Determination of cosmological parameters,
{\em Ap. J.} (2003) -accepted-
\bibitem{macro} M. Ambrosio et al. (MACRO Coll.),
 {\em Phys. Lett.} {\bf B434} (1998) 451; {\em Phys. Lett.} {\bf B357} (1995) 481.
\bibitem{miri} M. Ambrosio et al. (MACRO Coll.),
{\em Phys. Lett.} {\bf B566} (2003) 35.
\bibitem{sk} Y. Fukuda et al. (Super-Kamiokande Coll.), {\em Phys. Rev. Lett.}
{\bf 81} (1998) 1562; {\em Nucl. Instrum. Meth.} {\bf 503} (2003) 114.
\bibitem{bahcall2} J.N. Bahcall, M.H. Pinsonneault and S. Basu, {\em Ap. J.}
 {\bf 555} (2001) 990-1012.
 \bibitem{web} http://www.sns.ias.edu/\~{} jnb/
\bibitem{pdg} K. Hagiwara et al. (Particle Data Group), {\em Phys. Rev. } {\bf
D66} (2002) 010001.

\end{thebibliography}
\end{document}